\documentclass[fleqn,10pt]{wlscirep}
\usepackage[utf8]{inputenc}
\usepackage[T1]{fontenc}
\usepackage{bm}
\usepackage{glossaries}
\usepackage{tcolorbox}

\makeglossaries

\newglossaryentry{ordinary differential equations}{
  name=ordinary differential equations,
  description={An ordinary differential equation (ODE) is an equation which consists of one or more functions of one independent variable along with their derivatives. A differential equation is an equation that contains a function with one or more derivatives. But in the case ODE, the word ordinary is used for derivative of the functions for the single independent variable. ODEs are used to model the rate of change of the state of a system and have applications across various scientific disciplines, including physics, biology, and engineering}
}

\newglossaryentry{stochastic processes}{
  name=stochastic processes,
  description={Processes that involve a sequence of random variables changing over time. Stochastic processes are used to model systems or phenomena that are inherently unpredictable and have applications in fields like finance, physics, and biology}
}

\newglossaryentry{path}{
  name=path,
  description={In the context of stochastic processes, a path refers to a realization of the process, representing a possible sequence of states or events that the process can take over time}
}

\newglossaryentry{numerical solver}{
  name=numerical solver,
  description={An algorithm that approximates solutions to mathematical problems, especially differential equations, where analytical solutions are not feasible. They employ iterative numerical methods to simulate complex systems, crucial in various scientific and engineering applications. Their accuracy and methodology are adapted to suit different equation types and computational needs
}
}

\newglossaryentry{gradient back-propagation}{
  name=gradient back-propagation,
  description={A fundamental algorithm in neural network training, gradient back-propagation computes the gradient of the loss function with respect to the weights of the network. This process involves a forward pass to calculate outputs and loss, followed by a backward pass to calculate gradients. These gradients are then used to adjust the weights, optimizing the network's performance. It is crucial for training deep neural networks, enabling the effective learning of complex patterns in data
}
}

\newglossaryentry{universal approximator}{
  name=universal approximator,
  description={A universal
approximator is a parameterized object capable of representing any possible
function in some parameter size limit. Common universal approximators in low
dimensions include Fourier or Chebyshev expansions, while common universal
approximators in high dimensions include neural networks}
}

\newglossaryentry{Brownian motion}{
  name=Brownian motion,
  description={A type of stochastic process that models random motion, often used to represent particle movement. Brownian motion is a central concept in the theory of stochastic processes and is characterized by continuous, nowhere differentiable paths}
}

\newglossaryentry{Ito calculus}{
  name=Itô calculus,
  description={A branch of mathematical analysis that deals with stochastic integration and differentiation. Itô calculus is fundamental in the study of stochastic differential equations and is widely used in various fields such as finance and physics}
}

\newglossaryentry{Stratonovich calculus}{
  name=Stratonovich calculus,
  description={A formulation of stochastic calculus, similar to Itô calculus, but uses a different definition of the stochastic integral. The Stratonovich integral is often used when the stochastic system is derived from a physical model}
}

\newglossaryentry{Lipschitz-continuous and bounded}{
  name=Lipschitz-continuous and bounded,
  description={A property of functions where there exists a constant such that the absolute difference in function values is bounded by this constant multiplied by the absolute difference in input values. This property ensures stability and well-behavedness in mathematical models}
}

\newglossaryentry{variational inference}{
  name=variational inference,
  description={A method in Bayesian inference for approximating probability densities. This technique is widely used in machine learning for fitting complex models to data, especially in scenarios where exact inference is computationally infeasible}
}

\newglossaryentry{adversarial methods}{
  name=adversarial methods,
  description={Techniques in machine learning, particularly in neural networks, where models are trained against adversaries to improve robustness and performance, often used in generative models and for enhancing model security}
}

\newglossaryentry{adjoint-method}{
  name=adjoint-method,
  description={A numerical method for efficiently computing the gradient of a function or operator in a numerical optimization problem. In the context of neural differential equation it used for memory-efficient gradient back-propagation through the solution}
}

\newglossaryentry{manifold hypothesis}{
  name=manifold hypothesis,
  description={A foundational concept in the field of machine learning, suggesting that high-dimensional data (such as images, text, or neural data) usually lie on lower-dimensional manifolds within the high-dimensional space. This implies that although the data might exist in a space with a large number of dimensions, the intrinsic dimensions that actually govern the data's structure are much fewer. This lower-dimensional subspace or manifold captures the essential features and relationships in the data, making it possible to simplify and understand complex datasets}
}

\newglossaryentry{epistemic uncertainty}{
  name=epistemic uncertainty,
  description={Refers to uncertainty stemming from a lack of knowledge or information about a system or environment. It is often associated with the limited data or incomplete understanding of the underlying mechanisms of the system. Epistemic uncertainty can, in principle, be reduced through additional data collection, research, or analysis}
}

\newglossaryentry{aleatoric uncertainty}{
  name=aleatoric uncertainty,
  description={Pertains to uncertainty inherent to a system due to variability or randomness that cannot be reduced with more information. This type of uncertainty is intrinsic to the process or environment being observed and is often represented probabilistically}
}

\newglossaryentry{amortized priors}{
  name=amortized priors,
  description={In the context of machine learning and statistics, amortized priors refer to the concept where (some of) the parameters of the prior distribution are estimated across multiple observations or data points, rather than being entirely pre-defined. This approach allows for more efficient computation and learning, as the prior information is 'amortized' or spread out over the entire dataset, enabling the model to leverage shared information across different data instances}
}

 \title{Universal Differential Equations as a\\ Common Modeling Language for Neuroscience}

\author[1,*]{Ahmed El-Gazzar}
\author[1]{Marcel van Gerven}
\affil[1]{Donders Institute for Brain, Cognition and Behaviour, Radboud University, Nijmegen, the Netherlands}

\affil[*]{e-mail: ahmed.elgazzar@donders.ru.nl}

\begin{abstract}
The unprecedented availability of large-scale datasets in neuroscience has spurred the exploration of artificial deep neural networks (DNNs) both as empirical tools and as models of natural neural systems. Their appeal lies in their ability to approximate arbitrary functions directly from observations, circumventing the need for cumbersome mechanistic modeling. However, without appropriate constraints, DNNs risk producing implausible models,  diminishing their scientific value. Moreover, the interpretability of DNNs poses a significant challenge, particularly with the adoption of more complex expressive architectures. In this perspective, we argue for universal differential equations (UDEs) as a unifying approach for model development and validation in neuroscience. UDEs view differential equations as parameterizable, differentiable mathematical objects that can be augmented and trained with scalable deep learning techniques. This synergy facilitates the integration of decades of extensive literature in calculus, numerical analysis, and neural modeling with emerging advancements in AI into a potent framework. We provide a primer on this burgeoning topic in scientific machine learning and demonstrate how UDEs fill in a critical gap between mechanistic, phenomenological, and data-driven models in neuroscience. We outline a flexible recipe for modeling neural systems with UDEs and discuss how they can offer principled solutions to inherent challenges across diverse neuroscience applications such as understanding neural computation, controlling neural systems, neural decoding, and normative modeling. 
\end{abstract}
\begin{document}

\flushbottom
\maketitle

\thispagestyle{empty}

\section*{Introduction}
As holds for all the natural sciences, modern neuroscience is a scientific discipline whose advancement is fueled by both theoretical and experimental research~\cite{Urai2022,Churchland2016}. From a theoretical standpoint, we have witnessed important developments, ranging from detailed mechanistic models of specific neural circuits~\cite{Kim2017a, izhikevich2008large, felleman1991distributed, bliss1993synaptic} to grand unified theories of brain function~\cite{van1998dynamical, Friston2009a, hawkins2021thousand, miller2001integrative}. At the same time, from an experimental standpoint, advances in neurotechnolgy are allowing us to measure~\cite{Steinmetz2021, urai2022large, machado2022multiregion} and manipulate~\cite{Deisseroth2015, lozano2019deep, blumberger2018effectiveness} the activity of many thousands of neurons at an unprecedented scale.\\

\noindent A critical question is how to effectively integrate theoretical and empirical insights to expand our grasp of neural mechanisms and advance practical applications. In this perspective paper, we argue for universal differential equations (UDEs) as a unifying approach for neuroscience~\cite{rackauckas2020universal}. UDEs embrace the dynamical systems perspective on neuroscience, where neural systems are viewed as dynamical systems whose flow (dynamics) can be described in terms of systems of differential equations (DEs)~\cite{izhikevich2007dynamical,favela2021dynamical, durstewitz2023reconstructing}. Unlike conventional DEs, UDEs can be (partly or fully) estimated from data by marrying dynamical systems theory with machine learning.  
This formulation allows the integration of a-priori knowledge about the system along with high-capacity function approximators to model complex systems in the absence of large-scale datasets.
Consequently, UDEs are rapidly garnering attention across scientific domains where the datasets are still relatively scarce, and mechanistic, theory-driven models are prevalent, yet fall short in accounting for data variance~\cite{rackauckas2020universal,alquraishi2021differentiable, lai2021structural, karniadakis2021physics}. Similarly in neuroscience, differential equations are ubiquitous, underpinning the majority of theoretical, biophysical, and phenomenological models~\cite{izhikevich2007dynamical}. And despite their advances, existing measurements tools only provide sparse and noisy representation of the underlying neural mechanisms, which require both appropriate numerical tools and expert knowledge to guide modeling. In our view, UDEs provide a unique opportunity to bridge different modeling techniques, spanning various biological and abstraction scales in a unified framework to propel both fundamental and applied neuroscience.\\

\noindent To motivate UDEs, we begin with a critique on the current landscape of data-driven dynamical systems in neuroscience, highlighting key applications, and challenges, culminating in the motivation for hybrid approaches. Next, we delve into the taxonomy of UDEs in the context of stochastic dynamical systems and show how these mathematical objects provide a spectrum of modeling techniques familiar to the neuroscientist spanning from traditional mechanistic white-box models to sophisticated black-box deep learning models. We provide a general recipe for domain-informed training of UDEs for neural system identification and examine the benefits of UDE-based models in emerging applications within the field. We conclude by discussing current challenges and potential future directions.
Through this discourse, we argue that UDEs, when augmented with modern machine learning techniques, can serve as the foundational building block for multi-scale modeling in neuroscience, establishing a common language for theory formation and model development.

\section{Data-driven dynamical systems in neuroscience}\label{sec1}
\subsection*{Dynamical systems in neuroscience}
A prevalent perspective in neuroscience is viewing the brain as a dynamical system, availing the comprehensive toolbox of dynamical systems theory (DST) to the field~\cite{van1998dynamical, izhikevich2007dynamical, deco2008dynamic, breakspear2017dynamic, favela2021dynamical}. DST enables the formalization of mechanistic models as systems of differential equations or iterative maps~\cite{hodgkin1952quantitative, fitzhugh1961impulses, izhikevich2003simple} and provides a framework to explain properties of neural systems using intuitive geometrical and topological representations~\cite{deco2012ongoing, khona2022attractor}. This view also opens the door to adopt established phenomenological models and tools used in statistical physics to understand neural dynamics~\cite{wilson1972excitatory, kuramoto1975self, buzsaki2004neuronal}. However, both mechanistic and phenomenological models suffer from limitations that hinder their utility at scale~\cite{ramezanian2022generative}. Developing mechanistic models can be laborious and often insufficiently detailed, while phenomenological models typically offer only abstract descriptions of neural processes.\\

\noindent The unprecedented availability of large-scale datasets in neuroscience has spurred the exploration of data-driven dynamical systems, propelling the field into the era of big data~\cite{Landhuis2017}. These data-driven methods minimize reliance on a-priori assumptions, instead leveraging the rich data available to guide model identification~\cite{bruntonbook}. By training these systems to reconstruct empirical observations, they can act as direct surrogates to the system of interest.   This attribute makes them especially appealing within neuroscience~\cite{brunton2019data}, a field wherein the systems in question are notoriously complex to model, a unifying theoretical framework is still nascent, and the existing measurement tools do not currently provide a comprehensive representation of the underlying mechanisms.
Consequently, data-driven dynamical systems, and specifically deep recurrent neural networks (RNNs) and their variants, are increasingly integrated into a variety of research areas within neuroscience. In systems and computational neuroscience, data-driven dynamical systems are becoming valuable research tools for probing the neural underpinnings of cognitive and behavioral functions~\cite{durstewitz2023reconstructing, vyas2020computation, Shenoy2013, sussillo2014neural, barak2017recurrent, schaeffer2022no, mante2013context}. In neural control engineering, they are used to develop optimal neurostimulation profiles for treating clinical conditions~\cite{tang2018colloquium, acharya2022brain, yang2018control, bolus2021state, rueckauer2023silico}. Similarly, in neural decoding, they are used for reconstructing natural stimuli from neural recordings~\cite{willett2023high, metzger2023high, anumanchipalli2019speech, zhang2019survey, livezey2021deep}, advancing brain-computer interface technologies. Their applications extend to clinical neuroscience, where they are used for bio-marker discovery of psychiatric disorders, patient stratification, and prognosis~\cite{bystritsky2012computational, roberts2017clinical1, roberts2017clinical2, durstewitz2021psychiatric}.

\subsection*{Challenges}
The shift towards data-driven methodologies in neuroscience introduces significant technical challenges. These range from data-centric challenges such as high dimensionality, partial observability, non-linearity, process and measurement noise, non-stationarity, and data scarcity, to modeling hurdles such as uncertainty quantification, non-identifiability, and interpretability issues~\cite{durstewitz2023reconstructing}. This landscape has resulted in a plethora of specialized technical advancements driven by distinct theoretical and practical frameworks~\cite{brunton2019data, hurwitz2021building, ramezanian2022generative}. A symptom of this status quo is the prevalent dichotomy between model expressivity and interpretability. As researchers opt for more expressive models to capture the intricacies of neural dynamics, they encounter interpretability challenges. This is further exacerbated by optimization challenges that arise either due to the models (e.g. exploding/vanishing gradients in RNNs) or the behavior of the system (e.g. chaos and non-stationarity), entailing highly technical solutions that further fragments neuroscientific practice.\\

\noindent Additionally, while the allure of utilizing unbiased expressive models is initially appealing, in the absence of large-scale curated datasets, eschewing prior knowledge often results in ill-posed problems and implausible solutions as highlighted in recent studies~\cite{kao2019considerations, alber2019integrating, su2017false}. In practical terms, this means that the models become prone to overfitting on spurious correlations and exhibit high sensitivity to design choices that are peripheral to the main task at hand, ultimately leading to issues in generalization and replication across datasets, tasks, and subjects~\cite{maheswaranathan2019universality, schaeffer2022no, hurwitz2021building, han2023system}.

\subsection*{New frontiers}

Neural differential equations (NDEs)~\cite{chen2018neural, kidger2022neural} have emerged as a powerful tool of choice to implement data-driven dynamical systems. NDEs represent an emerging family of continuous models that utilize neural networks to parameterize the vector fields of differential equations. This integration marries the expressive power of neural networks with the rigorous theoretical foundations established by decades of research in differential equations and dynamical systems theory.  While originally popularized as deep neural network models with continuous depth~\cite{chen2018neural}, recent advancements have burgeoned into a rich spectrum of continuous-time architectures rooted in dynamical systems theory~\cite{tzen2019neural, morrill2021neural, li2020fourier, jia2019neural, poli2019graph, kidger2020neural}. Recently, neural ordinary differential equations are being increasingly adopted in computational and systems neuroscience, showing improved performance compared to current approaches~\cite{kim2021inferring, sedler2022expressive, geenjaar2023learning, versteeg2023expressive}. While this is a promising sign, their current application have only focused on black-box, explicitly discretized versions that do not capture the broader potential of NDEs as a pathway towards a unified scientific modeling language~\cite{shen2023differentiable, alquraishi2021differentiable, wang2023scientific}. This untapped potential can be realized by conceptualizing differential equations as parameterizable, differentiable mathematical objects amenable to augmentation and training via scalable machine learning techniques. Traditional DEs and NDEs can thus be viewed as special cases at the extreme ends of a spectrum.

\section{Universal differential equations}\label{sec2}
\begin{figure}[t]%
	\centering
	\includegraphics[width=\textwidth]{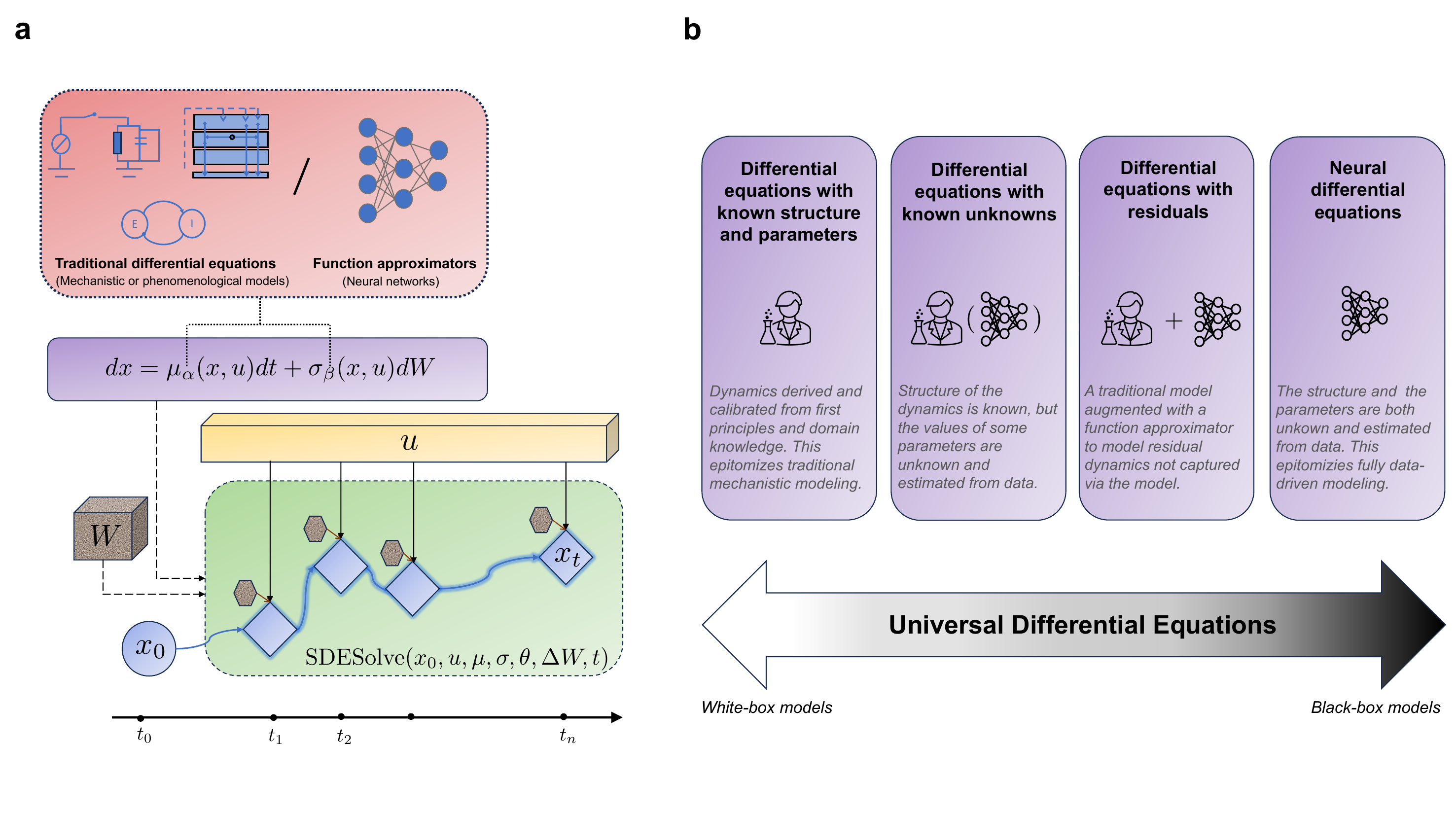}
	\caption{\textbf{Universal differential equations} a) A schematic illustration of a universal differential equation. The vector field of the differential equation is defined via either an existing model from the literature, or a differentiable universal approximator (e.g. a neural network) or a combination of both. The numerical solver is an SDE-compatible numerical solver, which takes in the initial condition $x_0$, a geometric Brownian motion generator $\Delta W$,  the forcing signal $u$, the functions defining the vector fields of the SDE $\mu$ and $\sigma$, along with their parameters $\theta$. The numerical solver then computes the solution at time $t$. The parameters of the differential equation can then be trained either via automatic differentiation or using adjoints methods. This setup enables the use of a UDE either as universal function approximator on their own or as a part (layer) in a differentiable computational graph. b) The formulation of a UDE encompasses a spectrum of modeling techniques from white-box traditional models to fully data-driven black box models. This flexibility can foster interoperability between different methodological efforts, provide solid theoretical background to face multifaceted challenges in neuroscience modelling across scales and applications, and offer a principled approach to balance between data adaptability and scientific rationale in model development.}\label{fig1}
\end{figure}

\subsection{Mathematical formulation}

A UDE is a mathematical model that extends a traditional differential equation by incorporating free parameters whose values can be learnt from data. By including free parameters, a UDE can act as a \gls{universal approximator}\cite{cybenko1989approximation, hornik1989multilayer}, meaning that it is able to approximate any dynamical system. 
In their most general form, UDEs are parameterized forced stochastic delay partial differential equations~\cite{rackauckas2020universal}. In this paper, we focus our attention on parameterized forced stochastic differential equations (SDEs).
SDEs extend \gls{ordinary differential equations} by incorporating \gls{stochastic processes}, enabling the modeling of dynamical systems subject to uncertainty. The key to this extension is the inclusion of a stochastic term that represents random fluctuations arising from either intrinsic or extrinsic factors. 
A forced SDE makes explicit how the (multi-dimensional) state $x(t)$ of a system of interest changes as a function of control inputs $u(t)$ and (Brownian) process noise $W(t)$ with $t$ the time index.  
This can be succinctly represented as
\begin{equation}\label{eq1}
    dx(t) = \mu_{\alpha}(x(t),u(t))dt + \sigma_{\beta}(x(t), u(t))dW(t) \,,
\end{equation}
where $\mu$ and $\sigma$ are drift and diffusion functions, representing the deterministic and stochastic parts of the time evolution of the system. Both $\mu$ and $\sigma$ are parameterized by $\theta = (\alpha, \beta)$, which are the (learnable) free parameters of the system.
Note that the time indices in Eq.~\ref{eq1} are typically suppressed from the notation for conciseness.\\

\noindent SDEs offer considerable flexibility for modeling stochastic dynamics. This adaptability largely stems from the diffusion term's configuration and Brownian motion properties~\cite{oksendal2013stochastic,sarkka2019applied}. For instance, in cases where $\sigma$ is a constant matrix or a state-independent function, the noise becomes \textit{additive}, rendering it suitable for modeling extrinsic uncertainties such as external, unobserved interactions. Conversely, when $\sigma$ is a function of the system's state, the noise becomes \textit{multiplicative}, which varies with the system's state, aptly capturing intrinsic uncertainties, such as uncertainties in drift term parameters. Notably, despite the Brownian noise process capturing uncorrelated Gaussian white noise, its interaction with $\sigma$ enables modeling of non-Gaussian noise distributions. These nuances provide a comprehensive framework for modeling complex dynamical systems with varying types of uncertainty. It is at the modeler's discretion to define the functional form of $\mu$ and $\sigma$. Ultimately, this functional form should accurately capture the (uncertain) evolution of the state of the system. This is evaluated by computing the solution to Eq.~\ref{eq1}, which is a distribution over paths $x(t)$ within some range $t \in [0, T]$. When this functional form is unknown, a feed-forward neural network becomes a conventional choice due to their ability to approximate any function.  Figure~\ref{fig3}a visualizes the temporal dynamics of a controlled SDE and Box 1 provides details on SDE solvers.

\subsection{Fitting a UDE to data}
The key idea behind efficient and scalable training of UDEs is the incorporation of a \gls{numerical solver} within a differentiable computational graph (Fig.~\ref{fig1}a). This setup allows \gls{gradient back-propagation} through the solution of the differential equation, enabling fitting the UDE parameters to observed data given a cost function. There are two primary strategies for this purpose: i) \textit{discretize-then-optimize}, which involves storing and gradient-backpropgation through all intermediate steps of the solver, providing exact gradients and ii) \textit{optimize-then-discretize}, utilizing the (stochastic) \gls{adjoint-method} to approximate gradients at fixed memory cost\footnote{Other methods, such as reversible solvers or forward sensitivity methods, also exist but are less common. For a detailed comparison of NDE automatic differentiation techniques, refer to~\cite{ma2021comparison} and ~\cite{kidger2022neural}.}. Effectively, this setup enables the training of a UDE-based model using standard loss functions similar to those used in discrete deep learning models. Nonetheless, given the stochastic nature of a UDE, maximizing the log-likelihood of observations alone can cause the diffusion function to converge to zero~\cite{li2020scalable}. To counteract this, alternative training strategies, including \gls{adversarial methods}~\cite{kidger2021neural} or \gls{variational inference}~\cite{li2020scalable, tzen2019neural, tzen2019theoretical}, are employed for stochastic UDEs. We explore in further detail the application of variational inference for UDE-based models in Section~\ref{sec3} and the technical details are provided in (Box 2).

\subsection{A continuum of models}

The UDE formulation naturally encompasses a spectrum of modeling approaches from traditional white-box mechanistic models to contemporary expressive black-box deep learning models (Fig. \ref{fig1}b).
Several modeling scenarios can thus be phrased as a UDE training problem. Here we provide some examples of these scenarios, where we use a subscript $\theta$ to indicate free parameters.

\subsubsection*{Differential equations with known unknowns}
In this setup, the structure of the system dynamics is known or assumed but the values of some parameters are unknown.
Training a UDE thus amounts to estimating these unknown parameters from observations.
This approach provide a structured and interpretable yet adaptable approach to modelling, capitalizing on domain knowledge or assumptions about the dynamics. This significantly reduces the model search space, and, if correct, would consequently reduce the amount of training required to approximate the dynamical system~\cite{linial2021generative, djeumou2023learn, abrevaya2023effective}.\\

\noindent Consider the following Ornstein-Uhlenbeck (OU) process used as a mechanistic model of the dynamics of a neuron's membrane potential~\cite{laing2009stochastic}:
\begin{equation}
    dx= a (m - x)dt + b dW \,,
\end{equation}
where $x$ denotes the membrane potential and $\theta=(a, m, b)$ are the free parameters. Here, $a$ indicates the rate of potential reversion to the mean, 
$m$ represents the resting membrane potential, 
and $b$ is the magnitude of random fluctuations due to synaptic inputs. Here the OU process provides the structure of the model dynamics, while the values of the parameters $\theta$ are estimated by fitting the UDE on empirical observations.

\subsubsection*{Differential equations with learnable uncertainty}
In this setup, the structure of the deterministic dynamics is known or assumed, with unknown parameters, and a function approximator is used to capture intrinsic and/or extrinsic uncertainty about the model.\\

\noindent Consider the modern interpretation of a Wilson-Cowan model~\cite{wilson1972excitatory}, used to describe the the average firing rates of a group of neurons~\cite{sussillo2014neural}. This model can be phrased as a UDE to capture stochastic dynamics not captured by the original model as follows:
\begin{equation}
        dx = \frac{1}{\tau}(-x + Jr(x)+ Bu)dt + 
        \sigma_{\beta}(x,u)dW \,,
\end{equation}
where $x$ represents the neurons' synaptic currents and $\theta=(\tau, J, B, \beta)$ are the free parameters. The function $r$ is a saturating nonlinear function and $J$ is a matrix that represents the recurrent connections between neurons. The vector $u$ represent the external input to the network that enters the system through the matrix $B$, and $\tau$ represents the time scale of the network. The function $\sigma$ is a differentiable function approximator (a neural network) that captures both how the dynamics respond to external unobserved inputs (extrinsic uncertainty) and how the dynamics evolve subject to uncertainty about the model parameters (intrinsic uncertainty). Hence, $\theta$ denotes the  parameters of the traditional model and the function approximator. These parameters are jointly learned by fitting the UDE on observations. 
This setup allows leveraging interpretable mechanistic deterministic models while embracing the complex stochastic nature that arise empirically when modeling complex systems from partial or noisy observations.

\subsubsection*{Differential equation with residual learning}

This approach assumes the knowledge a traditional model but part of the structure of the deterministic dynamics is unknown and could be captured via a universal function approximator. This is also often called residual dynamics learning. This enables generalizing powerful mechanistic models to handle rough edges not captured via the model or fill in missing information that is unknown or abstracted by the original.\\

\noindent For instance, consider the original Kuramoto model~\cite{kuramoto1975self}, widely used in neuroscience to study synchronization phenomena in systems of coupled oscillators (e.g. neurons, brain regions). A notable shortcoming of this model is its assumption of oscillator homogeneity, implying uniformity across all neurons or regions. However, biological systems often exhibit significant heterogeneity in terms of cell types, regional characteristics, and unobserved inputs. To accommodate these disparities, the Kuramoto model can be augmented  with a function approximator, allowing for a more precise representation of neural oscillations. This can be phrased as a UDE
\begin{equation}
	dx = \left( \omega + \frac{K}{N} \sum_{j=1}^N \sin(x_j - x) + f_{\alpha}(x) \right) dt + \Sigma\, dW \, ,
\end{equation}
where $x$ is a vector representing the phase of $N$ oscillators and $\theta=(\omega, K, \Sigma, \alpha)$ the free parameters with $\omega$ the natural frequencies, $K$ a matrix representing the coupling strength between the oscillators, and $\Sigma$ representing the magnitude of extrinsic random forces acting on the network. The function $f$, parameterized by $\alpha$, acts as a dynamic corrective mechanism, adjusting for deterministic discrepancies not accounted for in the original model formulation. Note that in this example, all the model parameters are assumed to be estimated from data.

\subsubsection*{Structured neural differential equations}

This setups posits that while the coupling architecture between the states (and inputs) of a system is known, the specific dynamic functions governing these states remain unidentified and can thus be approximated via a neural network. This approach is particularly apt for modelling complex, multi-scale, or networked non-linear dynamical systems.\\

\noindent Consider the following graph-coupled nonlinear dynamical system generalizing the Kuramoto model, phrased as a UDE:
\begin{equation}
	dx = \left( f_{\alpha}(x, u) + A \sum_{j=1}^N g_{\alpha}(x_j, x) \right) dt + \Sigma \, dW \,,
\end{equation}
where $x$ and $u$ represent the states and inputs of the system, respectively, and $\theta = (\alpha, \Sigma)$ are the free parameters. The functions $f$ and $g$ are function approximators describing the local and interconnected system dynamics, respectively. The matrix $A$ denotes the adjacency matrix representing the coupling structure, and $\Sigma$ represents the magnitude of extrinsic random forces acting on the network. In this particular example, the coupling structure is known or assumed based on a-priori assumptions (e.g. structural/functional connectivity), and the local and global dynamics function are completely learned from observations. In essence, either $f$ or $g$ could be replaced by traditional models allowing combining data-driven and mechanistic/phenomenological modeling across scales.

\subsubsection*{Neural differential equations}
In this configuration, both the parameters and structure of the system's dynamics are unknown. Consequently, the drift and diffusion vector fields of the UDE are entirely described via neural networks as function approximators. This setup represents the epitome of black-box modeling, as it enables the derivation of models directly from observational data circumventing the need for any assumptions about the system's dynamics. A generic UDE in this case can be written in the form of Eq.~\ref{eq1} as 
\begin{equation}
dx = \mu_{\alpha}(x,u)dt + \sigma_{\beta}(x,u)dW \,,
\end{equation}
where $\mu$ and $\sigma$ are neural networks with parameters $\theta = (\alpha, \beta)$. This equation can be viewed as a stochastic, continuous-time generalization of discrete-time deep recurrent neural networks prevalent in contemporary machine learning research~\cite{kidger2022neural, tzen2019theoretical, tzen2019neural, li2020scalable}.\\

\subsection{Towards informed stochastic models}
Generally, all of the presented UDE configurations fill a spectrum between white-box and black-box models under a unified formulation.  Intuitively, as one progresses from white-box models towards black-box models, the reliance on empirical data for model identification increases correspondingly, inversely proportional to the number of presupposed assumptions about the underlying dynamics (the more correct the model, the less data needed, and vice versa). In practice, it should be expected that a certain degree of knowledge or hypothesis about the studied system is available. This knowledge should not be constrained to the structure of the dynamics, but could  cover all aspects of the computational model (e.g., dimensionality, information about the stimulus or observation modality, scale of noise, expected dynamics, etc.). UDEs simply serve as a universal tool for evaluating this knowledge, or augmenting them to develop scalable models that can be used in several downstream applications (see Section~\ref{sec4}).\\

\noindent Crucially, UDEs conceptualize neural processes as continuous-time stochastic processes. This perspective can bring computational models closer to the complex nature of neural processes. This is imperative when modeling neural dynamics, where stochasticity can be traced from the molecular level, with stochastic behaviors in ion channels and synaptic transmission~\cite{hille1978ionic, sakmann2013single}, to the cellular scale where neurons demonstrate unpredictable firing patterns~\cite{tuckwell1988introduction}. Importantly, stochasticity is not confined to the micro-scale as it escalates to the level of neural populations, where the effects of noise and randomness are not merely incidental but play a crucial role in the functioning and organization of neural systems~\cite{rolls2010noisy, faisal2008noise}. The following section delves into leveraging UDEs to develop fully differentiable, informed, probabilistic models for neural system identification.

\section{Neural system identification}\label{sec3}

Let us consider a neural system whose dynamics are the realization of a continuous-time stochastic process 
$\{x(t) \colon 0 \leq t \leq T\}$
that is potentially modulated by exogenous input $u(t)$. 
In practice, we observe neither $x$ nor $u$ directly but rather have access to stimuli
$v_n = v(t_n)$ and neurobehavioural recordings $y_n = y(t_n)$, sampled at discrete timepoints $t_1,\ldots, t_N$ with $0 \leq t_n \leq T$. We also use $v_{1:N}$ and $y_{1:N}$ to denote these observations across timepoints. Let $\tau = (v_1,y_1,\ldots,v_N, y_N)$ denote a trajectory of stimuli and responses and assume that we have access to a dataset $\mathcal{D} = \{ \tau^1,\ldots, \tau^K \}$ consisting of $k$  such trajectories. The goal of neural system identification is to estimate the neural dynamics $x(t)$ from data $\mathcal{D}$.\\


\noindent We propose to model the underlying stochastic process $x$ as the (weak) solution of a latent UDE, and frame the problem of system identification as a posterior inference problem of the distribution $p(x \mid y, v)$, which we tackle via variational inference. Accurate resolution of this problem yields multiple benefits. First, it allows inference of the latent (hidden) states of the system in online and offline settings. Second, it allows reconstructing and predicting the system's behavior under various conditions. Third, it provides an expressive probabilistic modeling framework that quantifies uncertainty and incorporates prior knowledge, facilitating robust hypothesis generation and testing.\\

\noindent Recent methodological advancements in variational inference for SDE-based models are unlocking new avenues for probabilistic modeling of stochastic dynamical systems~\cite{li2020scalable, tzen2019neural, tzen2019theoretical, ryder2018black, course2023state}. This progress can offer transformative potential for neuroscience, specifically for modeling neural systems during complex naturalistic behavior. Here we outline an intuitive recipe, leveraging these techniques in conjunction with neuroscientific domain knowledge, aiming for an informed and data-efficient framework suitable to various applications in the field. We structure this recipe into four key modules, that is, (i) a stimulus encoder, (ii) a recognition model, (iii) a process model, and (iv) an observation model.

\begin{figure}[!t]%
\centering
\includegraphics[width=\textwidth]{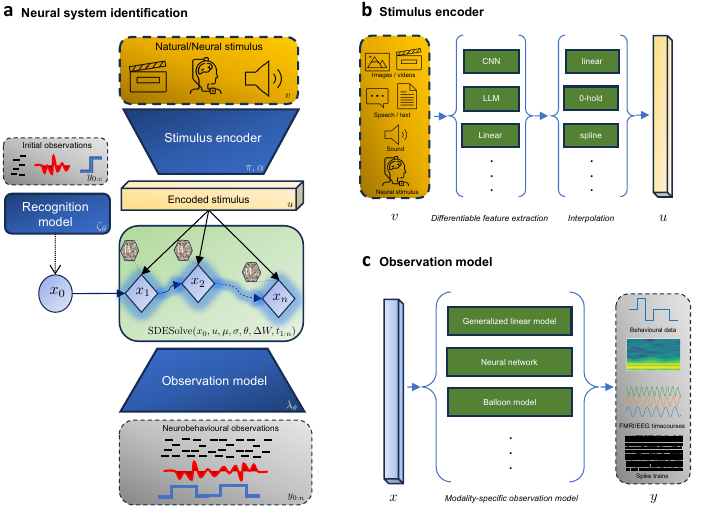}
\caption{\textbf{Framework for neural system identification} a) Shows the forward pass (generative mode) during the encoding of a (high-dimensional) stimulus $v$ into neurobehavioral observations $y$. This is done through a fully differentiable graph, which consists of i) a stimulus encoder to encode the stimulus into a lower dimensional continuous representation, ii) a recognition model to infer the hidden initial state $x_0$, iii) a latent dynamics model to model the temporal evolution of the dynamics, and iv) an observation model to  map the latent states into observations. b) Illustrates the formulation of the informed stimulus encoder which is tasked with learning a lower dimensional continuous representation $u$ from the discrete (high-dimensional) stimulus signal $v$. c) Illustrates examples of modality-specific observation models to map the latent process into neurobehavioral measurements.}\label{fig2}
\end{figure}

\subsection{Stimulus encoder}
The objective of this module is to map the  discrete-time measured stimulus $v$ into a (lower-dimensional) continuous-time  representation $u$ that is suitable for input into the latent dynamics. This is crucial in scenarios where the stimulus is a high-dimensional signal (e.g., images, videos, text, audio), as direct integration into the dynamics model would be computationally expensive. It is also crucial if we wish to sample the latent dynamics at a temporal resolution different from the temporal resolution of the measured input. We formalize this process as
\begin{equation}
u(t) = \pi \left\{ \alpha_\theta \left( v_{\tau} \right) \right\}_{\tau \leq t} \,,
\end{equation}
where \( \alpha_\theta \colon \mathbb{R}^{d_v} \to \mathbb{R}^{d_u} \) is a (parameterized) encoding function and \( \pi \colon \mathbb{R}^{d_u} \times [0,T] \to \mathbb{R}^{d_u} \) is the interpolation function that constructs a continuous representation over time. The design of both functions should be guided by the stimulus modality and the context of the scientific question being addressed.\\

\noindent The choice of interpolation scheme \( \pi \) should align with the temporal properties (e.g. smoothness, boundedness, missing data) of the stimulus~\cite{morrill2022choice, lepot2017interpolation}, the downstream application (e.g. online vs offline, speed vs accuracy), and the theoretical requirements of the drift term in the dynamics function (controlled differential equation~\cite{kidger2020neural} vs forced ODE). For example, while linear interpolation might suffice in most offline scenarios, if the model is to be used in real-time settings (e.g. control) then a rectilinear interpolation scheme is a suitable choice. A general recommended practice is to incorporate time as an additional input channel~\cite{kidger2022neural}, especially when the raw input lacks temporal variation, or to model non-autonomous dynamics.\\

\noindent 
The encoding function $\alpha$ can assume different forms, depending on the research question and data at hand. It could be a simple identity function in case the sensory input is low-dimensional. In case of high-dimensional input, leveraging one or more pre-trained models tailored to specific data modalities could offer a starting point (for instance, a pre-trained convolutional neural network for image data, or a pre-trained language model for text). Alternatively, \( \alpha \) may be parameterized by $\theta$ and learned directly from data.\\ 

\noindent Note that $\alpha$ could also be utilized to approximate the posterior distribution $p(u \mid v)$ instead of relying on point estimates which would fit well within variational inference framework. However, if the primary interest lies in parameterizing the underlying dynamical system, this added complexity may be unnecessary, as uncertainties about $u$ can also be captured through the diffusion term in the UDE. With that said, this approach could be more relevant in downstream applications such as neural decoding (see Section~\ref{sec4}).

\subsection{Recognition model}
The objective of this module is to accurately estimate the initial hidden state $x_0$ of the system. To accomplish this, we define a mapping function that uses a segment of the observed data to infer $x_0$ as suggested in recent studies~\cite{massaroli2020dissecting, rubanova2019latent}. This process involves a backward-running trainable sequential model, denoted as $\zeta_{\theta}$ (e.g. a RNN or neural controlled differential equation~\cite{kidger2021neural}). The recognition task can be expressed compactly as
\begin{equation}
    x_0 = \zeta_{\theta}\left(y_{c:0} , u_{c:0}\right) \,,
\end{equation}
where $c \in [0, N]$ denotes the end of the observation interval used for estimating the initial condition. Note that notations $y_{c:0} , u_{c:0}$ indicate that the intervals are reversed in time. The choice of $c$ will depend on the nature of the dynamics or context of the application. For example, in stationary settings, it might suffice to have $c \ll N$. It is also important to consider, which phenomena is under study. In most cognitive experiments, pre-task/stimulus recordings exist and can be utilized for this purpose. On the other hand, in online settings, dynamically sampling $c$ from a pre-defined distribution during training can adapt the model more effectively to real-time variations. In general it is important to ensure that $\zeta$  is not overly parameterized to avoid encoding future information about the dynamics as recommended by~\cite{massaroli2020dissecting}.
Additionally, it is worth noting that $\zeta$ can be employed to approximate the posterior distribution of the initial state $p(x_0 \mid y_{c:0},u_{c:0})$. However, one must consider the added complexity this introduces in the optimization process, especially when variational inference is to be applied across the entire dynamics of the system. Introducing this level of complexity might not always be necessary and could potentially complicate the model without significant benefits in certain contexts.. This approach could be relevant if the underlying dynamics of the system are assumed to be deterministic and autonomous, mirroring many current data-driven dynamical systems in neuroscience~\cite{chen2018neural, sedler2022expressive}.

\subsection{Process model}
The goal of this module is to learn the distribution of the latent stochastic process $x$. This is done by employing a UDE to model the temporal evolution of the initial state $x_0$, subject to external control $u$, and Brownian motion $W$. This is expressed as before as
\begin{equation}
dx = \mu_{\alpha}(x,u)dt + \sigma_{\beta}(x, u)dW \,.
\end{equation}
The design of the UDE should be dependent on domain knowledge about the system in question (Section~\ref{sec2}) and the downstream application of the model (Section~\ref{sec4}).
The parameters of the UDE can be trained along with the rest of the model via variational inference of $x$~\cite{li2020scalable, tzen2019theoretical, tzen2019neural} (Box 2).
For the reader familiar with conventional variational autoencoders (VAEs)~\cite{kingma2013auto}, it might be useful to conceptualize this as a variational autoencoder, conditioned on the stimulus~\cite{sohn2015learning}, with a (learned) expressive prior~\cite{ma2018constrained}, and whose latent space is an SDE-induced continuous stochastic process.

\subsection{Observation model}
Observation models, also known as measurement or emission models, define the probabilistic relationship between the latent states of a system and the observed data. The observation model is formalized as follows:
\begin{equation}\label{eq12}
    y(t) = \lambda_{\theta}(x(t), \epsilon(t)) \,,
\end{equation}
where $\lambda_
\theta \colon \mathbb{R}^{d_x} \to \mathbb{R}^{d_y}$ is the observation function and $\epsilon(t)$ is observation noise. The above mapping specifies the conditional distribution $p_{\theta}(y \mid x)$ within our probabilistic inference framework. The fidelity of observation models is paramount in the accurate identification of dynamical systems. These models must be tailored to reflect the biophysical constraints in the modality employed, account for specific noise structures, and possibly impose structure if the biological interpetability of latent states is required~\cite{klindt2017neural, seeliger2021end}. For instance, point process models can be employed for spike train data~\cite{heeger2000poisson}. Nonlinear Gaussian models may be apt for local field potentials (LFPs)\cite{herreras2016local}. Emission models for fMRI should account for the hemodynamic response function (HRF), possibly utilizing convolution or more complex models for regional variation~\cite{friston2000nonlinear}. Calcium imaging data necessitate nonlinear models to reflect the complex relationship between neural firing and observed signals, with considerations for photobleaching or other imaging artifacts~\cite{rahmati2016inferring}. 

\section{Opportunities in neuroscience}\label{sec4}
UDEs, when trained for neural system identification, offer the potential to serve as direct substitutes for various (data-driven) dynamical systems currently employed in neuroscience. Here we highlight four emerging applications within the field, outlining both the current challenges and the potential advantages of integrating UDE-based models over existing methodologies.

\begin{figure}[t]%
\centering
\includegraphics[width=\textwidth]{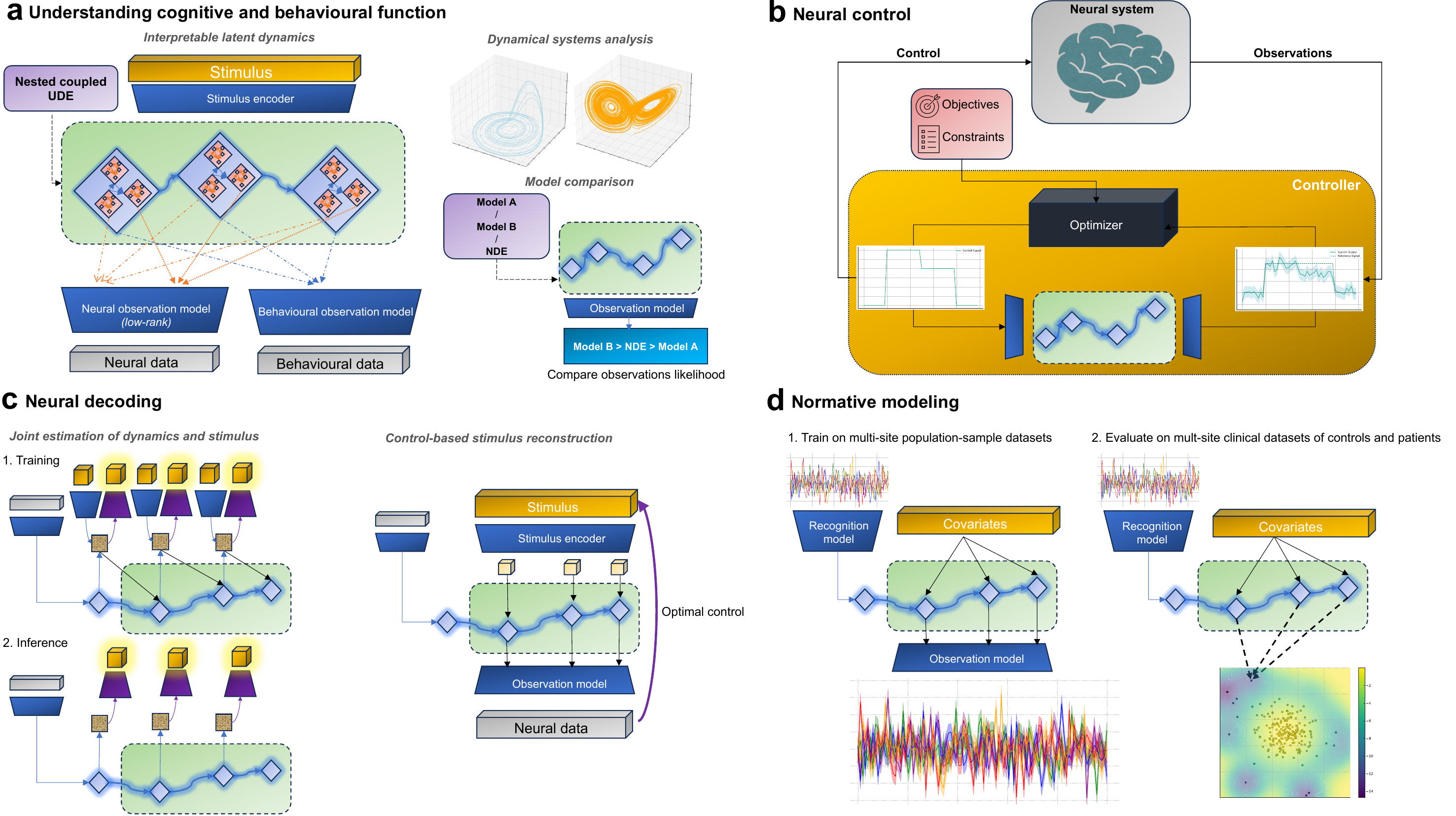}
\caption{\textbf{UDE-based models across different applications in neuroscience} \textbf{a:} UDEs can be used to represent the underlying latent dynamical system to understand how neural dynamics give rise to computations and ultimately behaviour. \textbf{b:} UDEs can be used for model-based closed loop control of neural systems. \textbf{c:} UDE models trained for neural encoding can also be leveraged for neural decoding of the stimulus from neural observations. \textbf{d:} UDE models can be employed for capturing population-average neural dynamics utilized for patients stratification in a normative modelling framework.}\label{fig3}
\end{figure}

\subsection{Explaining cognitive and behavioural functions}
A central question in neuroscience is how the brain implements cognitive and behavioural functions. In the past decade, significant progress have been gained by recognizing that these functions arise from coordinated activity within neural populations. This specific population-wide activity appears to be systematically governed by underlying lower-dimensional latent states~\cite{mante2013context,churchland2012neural,sadtler2014neural, elsayed2017structure}.
The characterization of latent state dynamics has been a primary focus of many latent variable models (LVMs)~\cite{yu2008gaussian, cunningham2014dimensionality, chen2018neural, glaser2020recurrent, kim2021inferring, hurwitz2021targeted, zhou2020learning, schneider2023learnable}. Despite their success in uncovering the neural basis of several phenomena, their utility is contingent upon several factors specific to the neuronal population and experimental settings at hand~\cite{urai2022large,hurwitz2021building}. These factors include low dimensionality of neural population activity, autonomous latent trajectories, and settings where most of the variance is explained by behaviour. Consequently, most breakthroughs in this domain have arisen from studies of highly stereotyped behaviors within simple contexts, particularly in neural populations in the motor cortex.
Additionally, several open challenges remain for current LVM approaches~\cite{vyas2020computation}, such as delineating the specific input-output structures of brain regions, understanding inter-neuronal population influences, modulating of local and global dynamics, and integrating both anatomical and functional constraints.\\

\noindent We posit that UDEs provide principled solutions to address these issues, making them a suitable candidate for multi-scale modelling of neural dynamics during complex behaviours. Note that LVMs, by design, will mix all sources of neural variability in the latent space, obscuring the interpretation of the latent dynamical system, especially for complex behaviors or diverse stimulus ensembles. UDEs offer a solution by imposing structured dynamics in the latent space, which can additionally improve model expressiveness and reduce the search space. The rich history of differential equations in neuroscience provides a foundation for such structure, which can be utilized to construct the drift term of a UDE and leverage the variational inference framework to infer their parameters. 
The effectiveness of this approach also crucially depends on the construction of appropriate encoder and observation models since this will affect the nature of inferred latent dynamics.\\

\noindent A particularly intriguing possibility that arises from the hybridization of mechanistic differential equations and neural networks in UDEs is developing expressive tractable multi-scale models. By imposing a-priori dynamics (with trainable parameters) at one scale while learning complex dynamics at another, we can combine different level of descriptions within a unified framework (Fig.~\ref{fig3}a). For example, existing single-neuron differential equations could model micro-scale dynamics, while neural networks could capture abstract inter-population macroscopic dynamics. Conversely, neural networks might model abstract microscopic dynamics within brain regions, and phenomenological models, guided by structural or functional insights, could represent macroscopic dynamics between brain regions. In this setup, UDEs function as a system of nested, coupled differential equations\cite{koch2023structural}. Again the ability to assign such correspondences would require the design of (low-rank) observation models that allow for the allocation of specific latent states to particular physical observations.\\

\noindent Another dimension of interpretability, which does not hinge on interpretable latent states, emerges from applying dynamical systems theory. By examining the vector field of the trained UDEs to identify dynamical phenomena and structures of interest (such as attractors, limit cycles, bifurcations, etc.), we can generate insights into how the dynamical system facilitates computations underlying cognition and behavior. This approach, drawing on principles from dynamical systems theory, has gained popularity in the analysis of RNN-based models~\cite{sussillo2013opening, durstewitz2023reconstructing, vyas2020computation}.\\

\noindent Contemporary methodologies for addressing stochasticity in neural dynamics often resort to simplified models of noise. Common approaches include using probabilistic initial states coupled with deterministic dynamics, or incorporating dynamics perturbed by  additive Gaussian or Poisson noise, as well as employing hidden Markov models~\cite{linderman2017bayesian, laing2009stochastic, pandarinath2018inferring}. While these methods can be adequate for modeling neural activity in autonomous tasks, discrete decision-making processes, or certain brain regions, they frequently fall short in encapsulating the full complexity inherent in higher-order brain functions or intricate decision-making scenarios. In such contexts, noise plays a more substantive role than just a source of randomness; it becomes a fundamental component of neural coding and behavior formation~\cite{rolls2010noisy, faisal2008noise}. UDEs represent a significant leap in the ability to model arbitrarily complex noise distributions. This is achieved through the integration of a high-capacity function approximator such as a neural network conditioned on the state in the diffusion term. The incorporation of such multiplicative noise, combined with the expressive capabilities of neural networks, allows for capturing  stochasticity arising from both unobserved interactions and intrinsic model properties. By embracing this stochastic nature, UDEs provide a more realistic and effective tool for understanding and simulating the dynamics of neural systems which might be overlooked by deterministic models or oversimplified noise representations.\\

\noindent UDEs offer a compelling framework for model comparison in neuroscience. Essentially, by leveraging the probabilistic inference framework, we can configure the prior UDE to reflect various theoretical models about how neural processes unfold over time. The crux of this approach lies in determining whether UDEs, when structured to reflect specific hypotheses, can improve the log-likelihood of observed data over models with non-specific or generic priors under identical training data conditions.
Such a comparison is not hypothesis testing in the formal Bayesian sense, but rather a way to evaluate the relative effectiveness of different (potentially highly expressive) dynamical models in explaining neural data, which otherwise would not be tractable in standard Bayesian settings~\cite{grimmer2011introduction}.\\

\noindent Beyond hypothesis testing, one of the most compelling aspects about adopting UDEs for neuroscience is the potential for automated scientific discovery.
This process is typically enabled by using sparsity-promoting optimization techniques~\cite{brunton2016discovering,schmidt2009distilling} to recover compact closed-form equations from a large database of basis functions. Within the framework of UDEs, this is viewed as a post-hoc step involving symbolic distillation of the function approximators to recover missing terms and auto-correct existing mechanistic models~\cite{cranmer2020discovering, rackauckas2020universal}. Unlike several scientific disciplines which are starting to embrace this approach~\cite{raissi2018deep, keith2021combining, davies2021advancing, duraisamy2019turbulence, kusner2017grammar, choudhary2022recent}, this remains an underexplored opportunity in neuroscience with a potential to generate data-driven hypotheses in the form of  interpretable algebraic expressions~\cite{wang2023scientific}.

\subsection{Neural control}
The confluence of neuroscience and control theory is becoming increasingly pronounced, spurred by the potential of brain-computer interfaces (BCI) and neurostimulation for clinical interventions, sensorimotor augmentation, and functional brain mapping\cite{yang2018control,acharya2022brain}. This burgeoning field, also termed "neural control engineering"~\cite{schiff2011neural}, is predicated on the notion that the brain is fundamentally a complex, adaptive system, amenable to modeling and control using established engineering and control theory principles. 
In practice, the predominant focus has been on open-loop control methods for neural systems. However, there is a growing consensus that transitioning to a closed-loop control paradigm is imperative to improve  reliability, safety and energy efficiency\cite{ramirez2018evolving, sarkka2019applied}. Particularly, model-based closed-loop control aids safety via allowing the development and validation of control strategies in-silico\cite{rueckauer2023silico}. It also promotes interpretability by facilitating causal analysis of the models~\cite{imbens2015causal}. Additionally, it paves the way for integrating the latest advancements at the intersection of control theory and machine learning in a data-efficient manner~\cite{moerland2023model}.\\

\noindent However, unlike the typical engineering context in which  model-based control methods are being developed, the brain posits a number of additional challenges~\cite{schiff2011neural}. These challenges encompass its high dimensionality, the multitude of constituent subsystems, limited data availability, inherent stochasticity, the myriad of spatio-temporal scales influencing system behavior, and technological constraints in sensing and actuation, among others. Consequently, there is a pressing need for tailored model-based frameworks specific to the brain~\cite{ramirez2018evolving}.\\

\noindent UDE-based models provide principled solutions to address these challenges, rendering them a strong candidate to develop models for neural control. The integration of mechanistic models with data-driven models in UDEs offer a way to navigate the current dichotomy of expressiveness versus data requirements. Fully mechanistic models, or linear data-driven models, while being less reliant on extensive data and potentially more interpretable, may compromise on prediction accuracy. Conversely, expressive non-linear data-driven dynamical systems can provide better prediction fidelity but necessitate substantial, often personalized, supervised datasets -- a requirement that can be challenging in practice.\\

\noindent One of the main concerns of closed-loop control system is its robustness to external disturbances and uncertainties. These uncertainties as discussed can come from all kind of sources in the neural system, ranging from \gls{epistemic uncertainty} about the system to \gls{aleatoric uncertainty} as generated by sensor, process and actuator noise. Such uncertainties is critical for safe and reliable design of control policies. A latent UDE model can provide a principled approach to estimate and disentangle these uncertainties even for highly expressive models in a tractable efficient manner. Latent UDEs facilitate the use of both linear and nonlinear control methods. Through the notion of \gls{amortized priors}, a distilled, simpler (possibly linear) model can be approximated from a complex dynamics model for real-time application, enabling optimal linear control strategies.  In applications where linear models are sub-optimal, we can still use an amortized non-linear UDE for system identification. The availability of a fully differentiable dynamics model then unlocks advanced control strategies, such as model-based reinforcement learning and gradient-based model predictive control~\cite{moerland2023model}. \\

\noindent For real-time applications, striking a balance between prediction accuracy and computational efficiency is paramount. UDEs, being compatible with adaptive numerical solvers, can thus be tailored to offer this trade-off crucial during real-time application via the choice of the numerical solver, or the trade-off between error tolerances and speed~\cite{yildiz2021continuous}. Additionally, their continuous-time nature allows for handling irregularly sampled heterogeneous data and guarantees adaptive continuous control in the absence of observations, ensuring safety of operation~\cite{lewis2012optimal}.

\subsection{Neural decoding}
Neural decoding utilizes activity recorded from the brain to make predictions about stimuli in the outside world ~\cite{rieke1999spikes, horikawa2013neural, anumanchipalli2019speech, seeliger2018generative, metzger2023high}. Such predictions can serve various purposes, from enabling communication interfaces and controlling robotic devices to improving our understanding of how brain regions interact with natural stimuli. As a result, neural decoding is rapidly becoming an indispensable framework in neuroscience~\cite{donoghue2002connecting, zhang2019survey}. Current approaches can be roughly broken down into two categories, in which the decoding algorithm is based on either regression techniques~\cite{warland1997decoding, horikawa2013neural, anumanchipalli2019speech, metzger2023high} or Bayesian methods~\cite{pillow2011model}.\\

\noindent Here, the architecture introduced earlier for system identification in Section~\ref{sec3} can be viewed as a neural encoding model. One option to reconfigure the architecture for neural decoding is to simply invert the input and output (and their corresponding encoder / decoder networks) during the training process to obtain a feasible trainable akin to modern supervised decoding models. A more powerful alternative is to utilize the same model for encoding, to also do decoding~\cite{paninski2007statistical, kriegeskorte2019interpreting}. Despite the ill-posed nature of the problem, we propose two approaches commonly used in modern control literature, utilizing the same architecture introduced earlier to potentially approach this problem in a tractable manner.\\

\noindent The first approach includes a modification to the architecture introduced earlier in Section~\ref{sec3} to extend the probabilistic inference framework to approximate the posterior distribution $p(v \mid y)$.  This involves updating the stimulus encoder to generate the parameters of an approximate posterior instead, a tractable prior over $p(u)$, and an additional decoder head to output the reconstructed stimulus (Fig.~\ref{fig3}c). We can then update optimization function accordingly. 
\noindent The second approach is to frame the problem of stimulus inference as a separate optimization problem akin to optimization problems solved in control applications. Recently, Schimel et al.~\cite{schimel2021ilqr} adopted this idea, by utilizing an iterative linear quadratic regulator (ILQR) within the recognition model of a sequential VAE. This method is used to estimate the initial state and infer any unobserved external inputs driving the system, demonstrating success on both synthetic and real-world neuroscience datasets. However, they note that the approach can be prone to local minima and may struggle with significant mismatches between the employed prior over the input and the posterior. They suggest that independently modeling process noise could mitigate these issues. This is an inherent advantage of latent UDEs, which naturally incorporate independent modeling of process noise, and could utilize this control-based approach for decoding.

\subsection{Normative modelling}
Normative modeling is an increasingly popular framework in clinical and developmental neuroscience that aims to characterize the \textit{normal} variation in brain features across a population and then assess individual deviations from this norm~\cite{marquand2016beyond, insel2010research, bethlehem2022brain}. This approach offers a statistical framework to correlate individual differences in brain metrics such as connectivity patterns, structural attributes, or task-induced responses with behavioural or clinical indicators. The appeal of normative modelling is becoming particularly pronounced in psychiatry. As the discipline increasingly recognizes the heterogeneity inherent in these measures~\cite{kapur2012has}, there is a concerted move towards eschewing symptom-based labels in favor of biologically grounded metrics. So far the emphasis in normative modelling has been on behavioural, structural neuroimaging, or static summaries of functional data ~\cite{wolfers2020individual, zabihi2019dissecting, rutherford2023evidence}. Developing normative models for dynamic representations of functional neuroimaging (e.g. EEG/fMRI timecourses) data remains a formidable challenge but necessary to characterize the majority of  psychiatric disorders~\cite{marquand2019conceptualizing, brodersen2014dissecting, rutherford2022normative, gazzar2022improving}. The challenge lies in capturing the high dimensional spatio-temporal dynamics of brain activity, which is further complicated by factors such as inter-subject variability, measurement noise (including physiological and scanner-related noise), and often limited sample sizes.\\

\noindent UDE-based models can provide a solution to navigate these challenges. Their flexible structure is adept at encapsulating a wide range of variability through the a complex diffusion term while capturing the population average behaviour in the drift term. The drift component of the UDE can further be parameterized with fixed arguments reflecting observed covariates within the population such as age, sex, scanning site, behavioural metrics, etc. At training time, the model can be optimized to reconstruct the neural data from observed variations and initial observations of population-sample or control only multi-site datasets~\cite{rutherford2022normative}. At test time, stratification is done via running the model on both control and patients and comparing their latent dynamics (or observations) log-likelihood (Fig.~\ref{fig3}d). This approach further provides a principled interpretable method to understand psychiatric disorders through the lens of network dynamics~\cite{durstewitz2021psychiatric, anyaeji2021quantitative, segal2023regional}.

\section*{Outlook and challenges}
There is a growing consensus that solutions to complex science and engineering problems require novel methodologies that are able to integrate traditional mechanistic modeling approaches and domain expertise with state-of-the-art machine learning and optimization techniques~\cite{raissi2019physics, alber2019integrating, willard2022integrating, cuomo2022scientific, alquraishi2021differentiable, de2019deep}. In this vein, we explore the potential of universal differential equations~\cite{rackauckas2020universal} as a framework to facilitate this integration in neuroscience. This endeavour is centered around the motivation of establishing a common potent language for modelling across the field.\\

\noindent In the realm of machine learning, differential equations and neural networks are increasingly being recognized as two sides of the same coin through the concept of neural differential equations~\cite{kidger2022neural}. For example, a residual neural network can be viewed as a discretized variant of a continuous-depth ODE with a neural network parametrizing its vector field~\cite{chen2018neural}.  Similarly, an RNN is equivalent to a neural controlled differential equation or a forced neural ODE, discretized via Euler approximation~\cite{kidger2020neural}. A convolutional neural network is roughly equivalent to the discretization of a parabolic PDE~\cite{li2020fourier, li2020neural}.\\
At first glance, these insights might not be entirely novel or exciting for a field such as computational neuroscience, which has been successfully applying RNNs and their variants as discretizations of ODE models for over a decade.
What is exciting however, is that this offers a fresh perspective to view and connect models from different scales of organization and levels of abstraction in neuroscience under one potent framework.\\

\noindent Beyond the advantages discussed throughout the rest of the paper, this perspective brings us closer to the language of dynamical system theory and classical differential equation literature. This alignment provides principled solutions to optimization and interpretation challenges in existing RNN-based models. For example, rough differential equations and log-ODEs might offer improved handling of long time series data~\cite{morrill2021neural}, partial differential equations (PDEs) create a natural bridge between dynamical systems and spatial domains~\cite{li2020fourier}, while SDEs offer structured ways to handle uncertainty~\cite{laing2009stochastic}. Being inherently continuous, these models adeptly handle irregularly sampled data and are compatible with powerful adaptive numerical solvers. All of these solutions could be explored and parsed through the formalism of UDEs.\\

\noindent Our focus in this paper is primarily on stochastic variants of UDEs, underscoring the empirical challenges in modeling neural systems. Despite advancements in neural recording technologies, the data obtained represents only a small, noisy subset of underlying mechanisms. Recent hypotheses suggest that behaviorally relevant neural dynamics may be confined to lower-dimensional spaces~\cite{mante2013context,churchland2012neural,sadtler2014neural, elsayed2017structure}. Yet, these representations may not consistently translate across time, task contexts, and brain regions. Challenges such as non-stationarity, intrinsic stochasticity of neural mechanisms, and the absence of a robust theoretical modeling framework raise doubts about our capacity to accurately model neural systems. Consequently, current models may over-rely on conditions where dynamics are autonomous or predominantly behavior-driven, as observed in cognitive experiments involving stereotyped tasks. These models are less effective in complex scenarios where the multi-scale spatial and temporal aspects of neural dynamics, such as in naturalistic behavior, become prominent. UDEs, as a form of SDEs, provide a framework to acknowledge and address these uncertainties by modeling neural processes as stochastic phenomena. Leveraging high-capacity function approximators in conjunction with SDE theory offers a pathway to navigate this challenging terrain. \\

\noindent We have presented a recipe for informed training of UDE-based models for neural system identification. This recipe leverages recent advancements in stochastic variational inference for SDE-based models and can be easily tailored to different downstream applications. While similar strategies are showing promising results across different applications ~\cite{course2023state, djeumou2023learn, fagin2023latent}, practical implementation in neuroscience is still warranted. Additionally, there remains several open practical questions and simplifying assumptions that warrant further research.\\

\noindent One assumption is modeling all uncertainties as standard Brownian motion within our dynamical systems. This perspective, while potent and aligning with the central limit theorem, can oversimplify real-world scenarios, where noise characteristics differ in bias and time-dependence. Recent advances in learning neural SDEs for fractional Brownian motion types offer avenues to better represent these complexities ~\cite{tong2022learning, daems2023variational}.\\

\noindent Secondly, the variational inference approach models the probability distributions of latent states and observations but makes point estimates for other inferred values, like initial states, model parameters, and encoded stimuli. While from a pragmatic point of view, this can be warranted,  especially given the nature of SDE-based dynamical systems, these practices could significantly influence the final model, and prevent formal Bayesian model comparison. With that said, recent developments in this area are promising and rapidly evolving. For example,~\cite{course2023state} provides an approach to jointly estimate the probability distribution of the SDE parameters using a reparametrization trick for Gaussian-Markov processes. This could be beneficial in our setup if the drift in the UDE is dependent on the the parameters of the mechanistic model alone and no neural networks are used. Alternatively,~\cite{zeng2023latent} provides an efficient approach for probabilistic inference of latent neural SDEs that operate on homogeneous manifolds, an assumption ubiquitous in neuroscience. These developments pave the way towards fully probabilistic treatment of our models.\\

\noindent Lastly, the field of neural differential equations is relatively nascent compared to established deep learning practice. Challenges remain, particularly in gradient back-propagation through numerical solvers. In some scenarios where the dynamics are stiff or discontinuous, training via automatic differentiation with high-order adaptive numerical solvers can be very expensive in terms of memory and speed~\cite{ma2021comparison}. Alternatively, training via adjoint-sensitivity methods can be more memory-efficient but still remains slow and results in biased gradients. Innovative solutions like algebraically reversible solvers~\cite{kidger2021efficient}, and stochastic automatic differentiation~\cite{arya2022automatic} are emerging, but their mainstream adoption is still in early stages.\\

\noindent With that said, the challenges discussed are not unique to neuroscience but resonate across various scientific disciplines. This burgeoning field of scientific machine learning is a collaborative and innovative arena, marked by rapid advancements. The recent surge in open-source software and packages~\cite{rackauckas2017differentialequations, rackauckas2019diffeqflux, lienen2022torchode, tuor2021neuromancer, kidger2021equinox} centred around neural and universal differential equations specifically, and automated model discovery in general, underscores the growing interest and recognition of a new era of scientific discovery~\cite{wang2023scientific}. Neuroscience is poised to embrace this new era, to push the boundaries of our current understanding of the brain and advance practical applications in the field. \\

\begin{tcolorbox}[title=Box 1: Solving stochastic differential equations, colback=white, colframe=blue, colbacktitle=lightgray, coltitle=black, fonttitle=\bfseries]
Consider the general form of an SDE
\begin{equation}
dx = \mu(x, t)dt + \sigma(x, t)dW \,,
\end{equation}
where $\mu$ and $\sigma$ represent the deterministic drift and stochastic diffusion components, respectively. This non-autonomous (time-varying) formulation also includes forced SDEs since we may use the substitutions $\mu(x(t), t) = \mu'(x(t), u(t))$ and 
$\sigma(x(t), t) = \sigma'(x(t), u(t))$.
The system's state \(x(t)\) in this case is a stochastic process, and can be obtained by solving the differential equation as follows
\begin{equation}
    x(t) = x(0) + \int_{0}^{t}\mu(x(s), s)ds + \int_{0}^{t}\sigma(x(s), s)dW(s) \,.
\end{equation}

Two main challenges arise in attempting to solve this equation. 
Firstly, Brownian motion is almost surely nowhere differentiable, rendering standard (Riemann) integration infeasible for the last term. This necessitates the application of stochastic calculus with one of two primary frameworks: \gls{Ito calculus}, or \gls{Stratonovich calculus}. While each has its typical applications and advantages, they coincide when $\sigma$ is a fixed matrix, and the choice becomes arbitrary when it is a learned function. 
Secondly, deriving analytical solutions for non-trivial SDEs is  typically infeasible. Thus we resort to numerical solvers compatible with SDEs~\cite{oksendal2013stochastic} (e.g Euler–Maruyama, Milstein, stochastic Runge–Kutta) to approximate the solution at discrete time points via
\begin{equation}\label{eq3}
x_{n+1} \approx {\tt SDESolve}(x_n, \mu, \sigma, \Delta w, \Delta t).
\end{equation}
for $0 \leq t_n \leq T$ with $\Delta w$ a Brownian increment and $\Delta t = t_{n+1} - t_n$ the time step.

SDEs have seen widespread application for modeling real-world phenomena, such as asset prices in finance~\cite{black1973pricing, cox2005theory}, particle dynamics in physics~\cite{coffey2012langevin}, population dynamics in ecology~\cite{arato2003famous}, and allele expression in genetics~\cite{ewens2004mathematical}. Historically, however, constructing the structure of SDEs has relied heavily on substantial domain expertise. Additionally, fitting the parameters of SDEs to data has been dependent on techniques that fail to scale favorably~\cite{yang1991monte, chow2015path, gobet2005sensitivity}. These limitations have often resulted in oversimplified models lacking the representational capacity to accurately capture the intricacies of complex processes.
The recent advent of universal differential equations heralds a new era of opportunities to develop expressive and scalable SDE models for modelling and estimating complex dynamical systems. 

\end{tcolorbox}

\begin{tcolorbox}[title=Box 2: Variational inference of latent UDE models, colback=white, colframe=blue, colbacktitle=lightgray, coltitle=black, fonttitle=\bfseries]

To perform variational inference in the context of latent continuous stochastic processes, we need to define a reasonable and tractable family of path distributions for both the prior and the approximate posterior. Following the approach of~\cite{li2020scalable}, we may employ two SDEs to represent these distributions. Specifically, here we can define our (amortized) prior as a UDE and our approximate posterior another black-box UDE. Note  that the term prior here refers to our main generative UDE, and the approximate posterior is an auxiliary UDE that is used only during training. The UDEs can be written as: 
\begin{align}\label{eq10}
    dx &= \mu_{\theta}(x,u)dt + \sigma_{\theta}(x,u)dW \,, \quad x(0)=x_0 \,, \quad \text{(prior)} \,, \\
    d\tilde{x} &= \mu_{\phi}(\tilde{x}, y, u)dt + \sigma_{\theta}(\tilde{x},u)dW \,, \quad \tilde{x}(0)=x_0 \,, \quad \text{(approx. posterior)} \,,
\end{align}
where $\phi$ are the variational parameters. Note that both processes share the same diffusion $\sigma_{\theta}$, a decision which guarantees that the Kullback-Leibler (KL) divergence between the two probability measures they induce is finite (under some mild conditions)~\cite{li2020scalable}. This KL divergence can be defined using Girsanov's theorem~\cite{girsanov1960transforming} as\begin{equation}\label{eq11}
\textrm{KL}(Q || P) = \mathbb{E}_{Q}\left[ \int_0^T \left( \frac{1}{2} \left\| \sigma_{\theta}(\tilde{x},u)^{-1} \left( \mu_{\phi}(\tilde{x}, y, u) - \mu_{\theta}(\tilde{x},u) \right) \right\|^2 \right) dt \right] \,,
\end{equation}
where $Q$ and $P$ denote the path space probability measures induced by the approximate posterior and prior UDEs, respectively. Intuitively, this KL divergence
resembles the integrated difference over the time horizon
$[0, T]$ between the prior drift $\mu_{\theta}$ and posterior drift $\mu_{\phi}$, scaled by the diffusion $\sigma_{\theta}$. This divergence can be estimated up to a constant with Monte Carlo, sampling trajectories from the dynamics given by the approximate posterior.

The neural system identification framework presented in Section~\ref{sec3} can be interpreted as a variational autoencoder~\cite{kingma2013auto}, conditioned on the stimulus~\cite{sohn2015learning}, with a (learned) expressive prior~\cite{ma2018constrained}, and whose latent space is an SDE-induced continuous stochastic process~\cite{li2020scalable}.
We can train the model parameters $\Theta = [\theta, \phi
]$ by maximizing the evidence lower bound (ELBO) formulated as follows:
\begin{equation}\label{eq13}
\text{ELBO}(\Theta; v, y) = \mathbb{E}_{Q}\left[\sum_{i=0}^{n}\text{log }p_{\theta}(y(t_i) \mid x(t_i),v(t_i))\right] - \textrm{KL}(Q || P) \,,
\end{equation}
where the first term represents the log-likelihood of the observations given the latent states and the stimulus, and the second term represents the KL divergence between the prior and approximate posterior.

With the optimization function in place, we can compute the gradients with respect to the model parameters $\Theta$ and backpropagate through our fully differentiable computational graph either via automatic differentiation through the solver operations or via the stochastic adjoint-method~\cite{li2020scalable}.

In essence, we can use the variational inference framework to provide an approximate Bayesian treatment of all the inferred values such as the initial state $x_0$, the encoded input $u$, and the UDE parameters $[\theta,\phi]$ ) and update Eq.~\eqref{eq13} accordingly. However, this comes at the cost of a more challenging optimization problem to solve. Therefore, pragmatic choices must be made based on the specific application requirements.
\label{Box2}
\end{tcolorbox}
\noindent\textbf{Acknowledgements}\\
This publication is part of the project Dutch Brain Interface Initiative (DBI$^2$) with project number 024.005.022 of the research programme Gravitation which is (partly) financed by the Dutch Research Council (NWO).

\noindent\textbf{Author contributions}\\
AE: Conceptualization, literature review, and manuscript writing.\\
MvG: Conceptualization, supervision, and manuscript revision.

\noindent\textbf{Competing interests}\\
The authors declare no competing interests. \\

\printglossaries

\bibliography{refs}
\end{document}